\documentclass[twocolumn,apjl]{openjournal}
\usepackage{graphicx}
\usepackage{amsmath}
\usepackage{amssymb}
\usepackage{xspace}
\usepackage[dvipsnames]{xcolor}
\usepackage[backref, pagebackref=false, hyperindex=true, breaklinks=true, colorlinks=true, urlcolor=magenta, linkcolor=NavyBlue, citecolor=cyan, citebordercolor={0 1 0}, pagecolor=red, bookmarks=true, filecolor=blue,bookmarksopen=true, plainpages=false, pdfpagemode=UseThumbs]{hyperref}
\hypersetup{
colorlinks=true,
linkcolor=blue,
filecolor=blue,
urlcolor=blue,
citecolor=blue,
}
\usepackage{multirow}
\usepackage{soul}
\usepackage{multirow}
\newcommand{\figref}[1]{Fig.~\ref{#1}}
\newcommand{\Figref}[1]{Figure~\ref{#1}}
\newcommand{\ITdd}{\texttt{ITdd}}

\newcommand{\fargo}{{\sc fargo}}
\newcommand{\fargotwodoned}{{\sc fargo-2d1d}}
\newcommand{\fargoadsg}{{\sc fargo-adsg}}
\newcommand{\fargothreed}{{\sc fargo3d}}
\newcommand{\idefix}{{\sc idefix}}
\newcommand{\phantomcode}{{\sc phantom}}
\newcommand{\sbt}{\,\begin{picture}(-1,1)(1,-3)\circle*{3}\end{picture}\ }

\begin{document}

\title{The reflex instability: \\exponential growth of a large-scale $m=1$ mode in astrophysical discs\vspace{-16pt}}

\shorttitle{The reflex instability}

\shortauthors{Crida et al.}

\author{Aurélien Crida}
\affiliation{Université Côte d’Azur, Observatoire de la Côte d’Azur, CNRS, laboratoire Lagrange UMR7293, Nice, France\vspace{-8pt}}

\author{Clément Baruteau}
\affiliation{IRAP, Université de Toulouse, CNRS, Université Paul Sabatier, CNES, Toulouse, France\vspace{-8pt}}

\author{Jean-François Gonzalez}
\affiliation{Universite Claude Bernard Lyon 1, CRAL UMR5574, ENS de Lyon, CNRS, Villeurbanne, F-69622, France\vspace{-8pt}}

\author{Frédéric Masset}
\affiliation{Instituto de Ciencias Físicas, Universidad Nacional Autonoma de México, Av. Universidad s/n, 62210 Cuernavaca, Mor., Mexico\vspace{-8pt}}

\author{Paul Segretain}
\affiliation{Université Côte d’Azur, Observatoire de la Côte d’Azur, CNRS, laboratoire Lagrange UMR7293, Nice, France\vspace{-8pt}}

\author{Philippine Griveaud}
\affiliation{Université Côte d’Azur, Observatoire de la Côte d’Azur, CNRS, laboratoire Lagrange UMR7293, Nice, France}
\affiliation{Max-Planck-Institut für Astronomie, Königstuhl 17, 69117 Heidelberg, Germany\vspace{-8pt}}

\author{Héloïse Méheut}
\affiliation{Université Côte d’Azur, Observatoire de la Côte d’Azur, CNRS, laboratoire Lagrange UMR7293, Nice, France\vspace{-8pt}}

\author{Elena Lega}
\affiliation{Université Côte d’Azur, Observatoire de la Côte d’Azur, CNRS, laboratoire Lagrange UMR7293, Nice, France}

\begin{abstract}
  We report the finding of a linear, non-axisymmetric, global instability in gas discs around stars, which may be relevant to other astrophysical discs. It takes the form of an $m=1$ mode that grows in the disc density distribution while the star-barycentre distance rises exponentially with a characteristic timescale that is orders of magnitude longer than the orbital period.
  We present results of hydrodynamical simulations with various codes and numerical methods, using either barycentric or stellocentric reference frames, with or without the disc's self gravity: all simulations consistently show an unstable mode growing exponentially.
  
  The instability disappears if, and only if, the reflex motion of the star due to the disc's asymmetry is not taken into account in the simulations. For this reason we refer to this instability as the \emph{reflex instability}.
  We identify a feedback loop as a possible origin, whereby the acceleration of the star excites the eccentricity of the disc, yielding an $m=1$ mode in the density distribution which, in turn, pulls the star.
  The growth timescale of the instability decreases with increasing disc mass and is a few hundred orbits for disc-to-star mass ratios of a few percent. If truly physical, and not due to a numerical artifact that would be common to all the codes we have employed, the reflex instability could have a dramatic impact on protoplanetary discs evolution and planetary formation.
\end{abstract}

\keywords{accretion, accretion discs --- hydrodynamics --- methods:
  numerical --- planetary systems: formation --- planetary systems:
  protoplanetary discs\vspace{12pt}}

\maketitle

% ================
\section{Introduction}
\label{sec:intro}
% ================

Gravitational systems in astrophysics often comprise a body -- the primary -- that far outweighs the others.
A typical example is protoplanetary discs (with or without planets) around single stars.
In this Letter, we show that such a situation leads to an instability which had so far not been reported (to our knowledge), probably because its growth time was too long for the simulations times and disc masses considered usually, or because the strange behaviour of the gas disc was attributed to some other phenomenon. However, when running simulations on thousands of orbits with disc-to-star mass ratios larger than $10^{-3}$, its appearance is inevitable. We argue that this instability is of physical, not numerical, origin, and occurs as soon as the disc's axisymmetry is broken.
The key ingredient is the reflex motion of the star in reaction to an asymmetry in the disc, often called the Indirect Term in stellocentric codes (and sometimes discarded). A complete discussion of the meaning, the role, and the relevance of the Indirect Terms in general is proposed in our companion paper \citep[][hereafter Paper~I]{Crida+2025a}. 

Our physical setup and the various codes we use are presented in Sect.~\ref{sec:model}, before 
the results of our simulations and a description of the instability are reported in Sect.~\ref{sec:overview}.
A physical interpretation is proposed and the role of the different ingredients is analysed in Sect.~\ref{sec:interpretation}. Finally we discuss the implications of our work and compare with other results in the literature in Sect.~\ref{sec:discussion}, notably in the field of N-body problems and galactic dynamics, where similar results have been found.

% ================
\section{Physical model and numerical setup}
\label{sec:model}
% ================

% ----------------------------
\subsection{Physical equations and reference values for the disc parameters}
% ----------------------------
The physical system that we consider is comprised of a protoplanetary gaseous disc orbiting around a star.
Our fiducial disc model is deliberately simple: razor-thin (two-dimensional), unmagnetized, inviscid/non-turbulent, isothermal.
The disc is described by polar coordinates ($R, \varphi$) and its evolution is modelled by the continuity equation, Euler's equation and the ideal gas law, with a temperature both uniform and stationary.
The choice of a globally isothermal disc allows to avoid the Vertical Shear Instability when examining the onset of our instability in supplementary 3D simulations \citep{Nelson+2013}.
The disc's self-gravity is included in some of our 2D simulations, which further solve Poisson's equation.

Our disc model assumes a uniform initial radial profile for the Toomre-Q parameter \citep[][see also footnote 7 in Paper~I]{Toomre1964}.
Since the disc is assumed to be isothermal, the aspect ratio $h(R) \propto R^{1/2}$ so a uniform $Q$ implies that the initial disc's surface density profile scales as $R^{-3/2}$.
The instability mechanism examined in this Letter being non-axisymmetric, white noise is added to the initial density profile at the 0.1\% level in order to break the initial symmetry.

Our simulations adopt the following code units: a reference radius $R_0$ as the length unit, the stellar mass $M_{\star}$ as the mass unit, and the inverse Keplerian frequency at $R_0$ as the time unit.
In the results shown in Sect.~\ref{sec:overview}, time is expressed in orbital periods at $R_0$: $T_0=2\pi(GM_\star/R_0^3)^{-1/2}$.
Our reference model has $\Sigma(R_0) = 4\times10^{-3}$ and $h(R_0) = 0.05$, thus $Q \sim 4$ initially.
The disc extends from 0.32$R_0$ to 3.125$R_0$, so our fiducial value for $\Sigma(R_0)$ implies an initial disc-to-star mass ratio of $\sim$0.06.
We stress that such a physical set-up should be stable according to the present common knowledge on Keplerian discs.

% ----------------------------
\subsection{Numerics}
% ----------------------------
We have used several codes to study the linear growth stage of the instability:
%
%FFFFFFFFF
\begin{figure*}[ht]
    \centering
    \includegraphics[width=0.33\hsize]{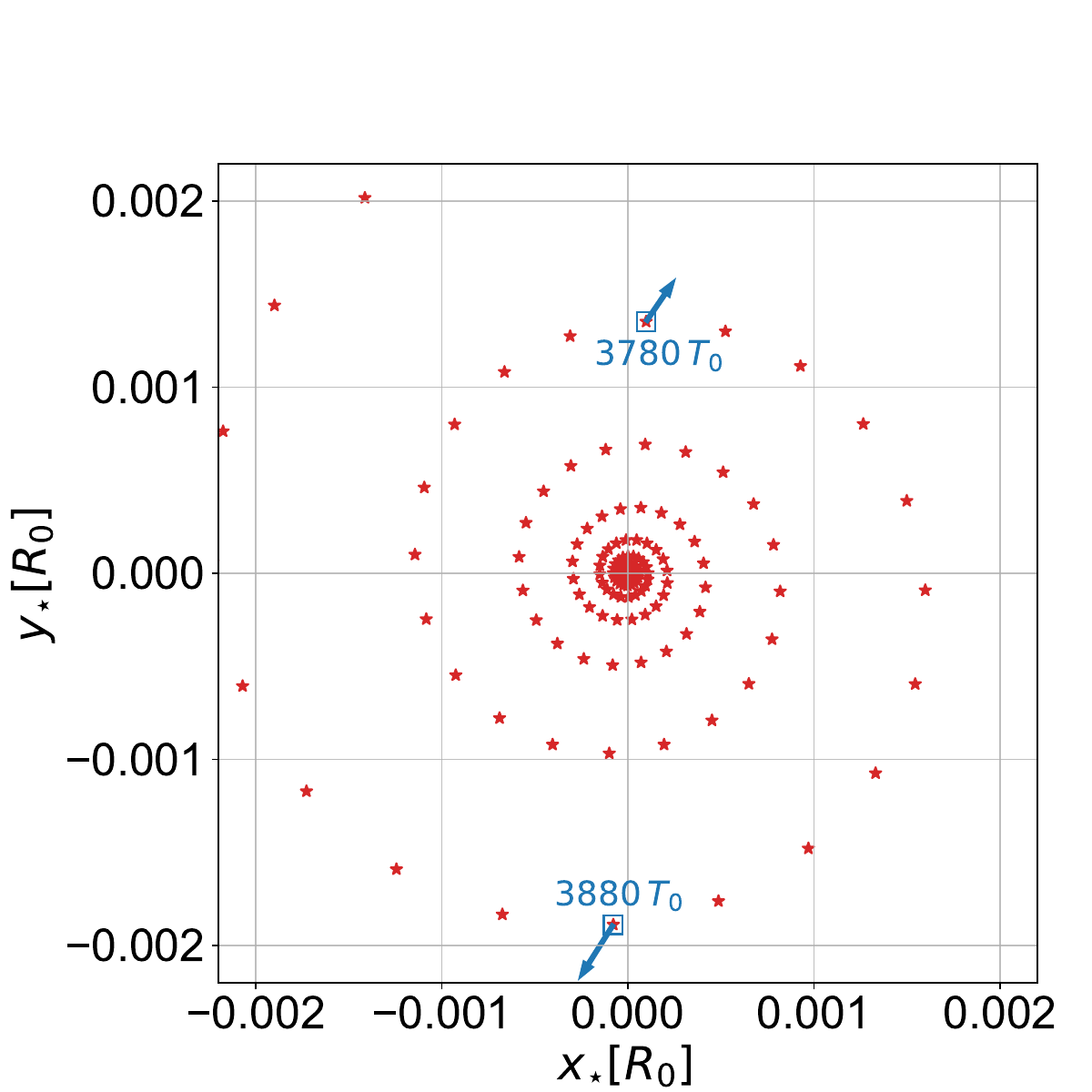}
    \includegraphics[width=0.33\hsize]{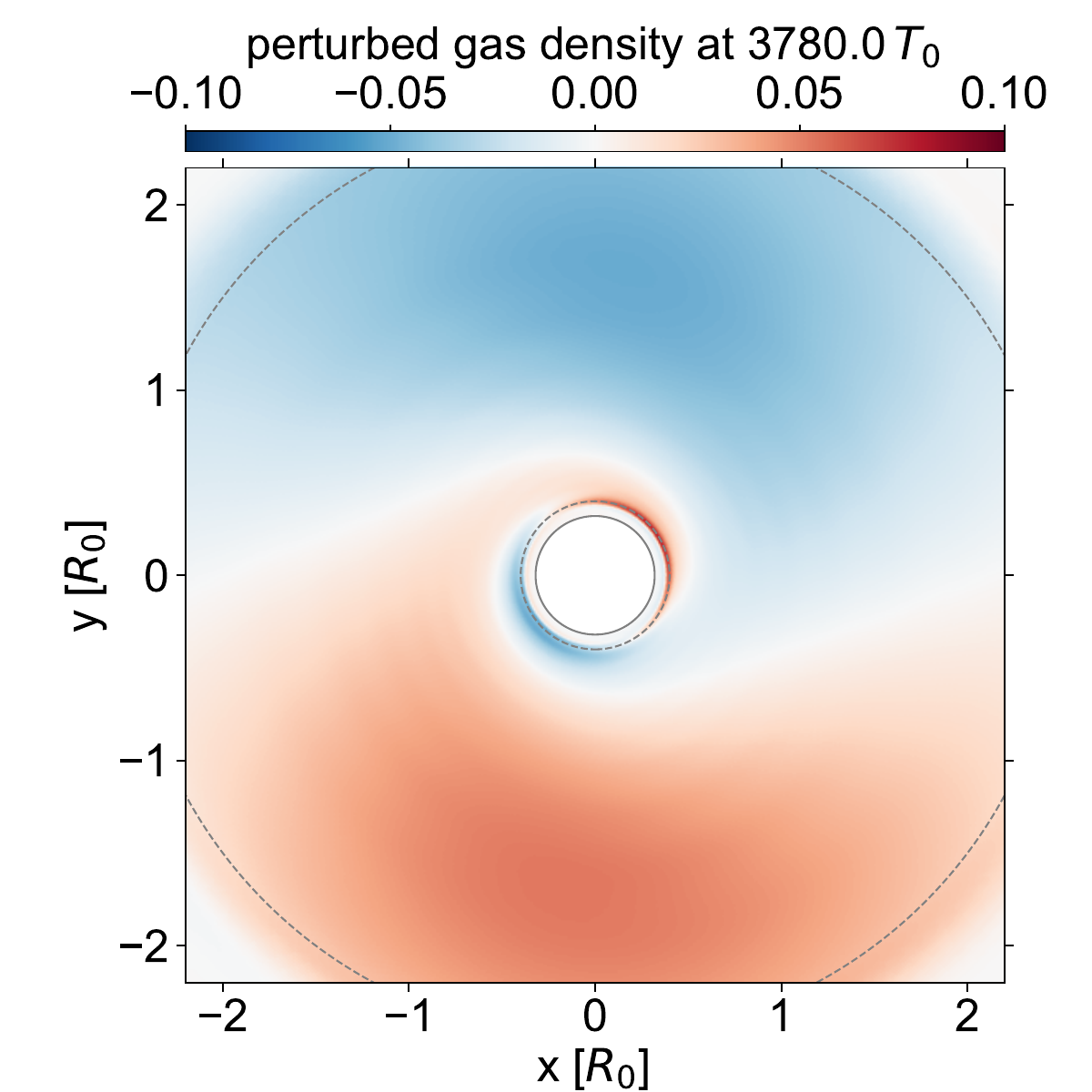}
    \includegraphics[width=0.33\hsize]{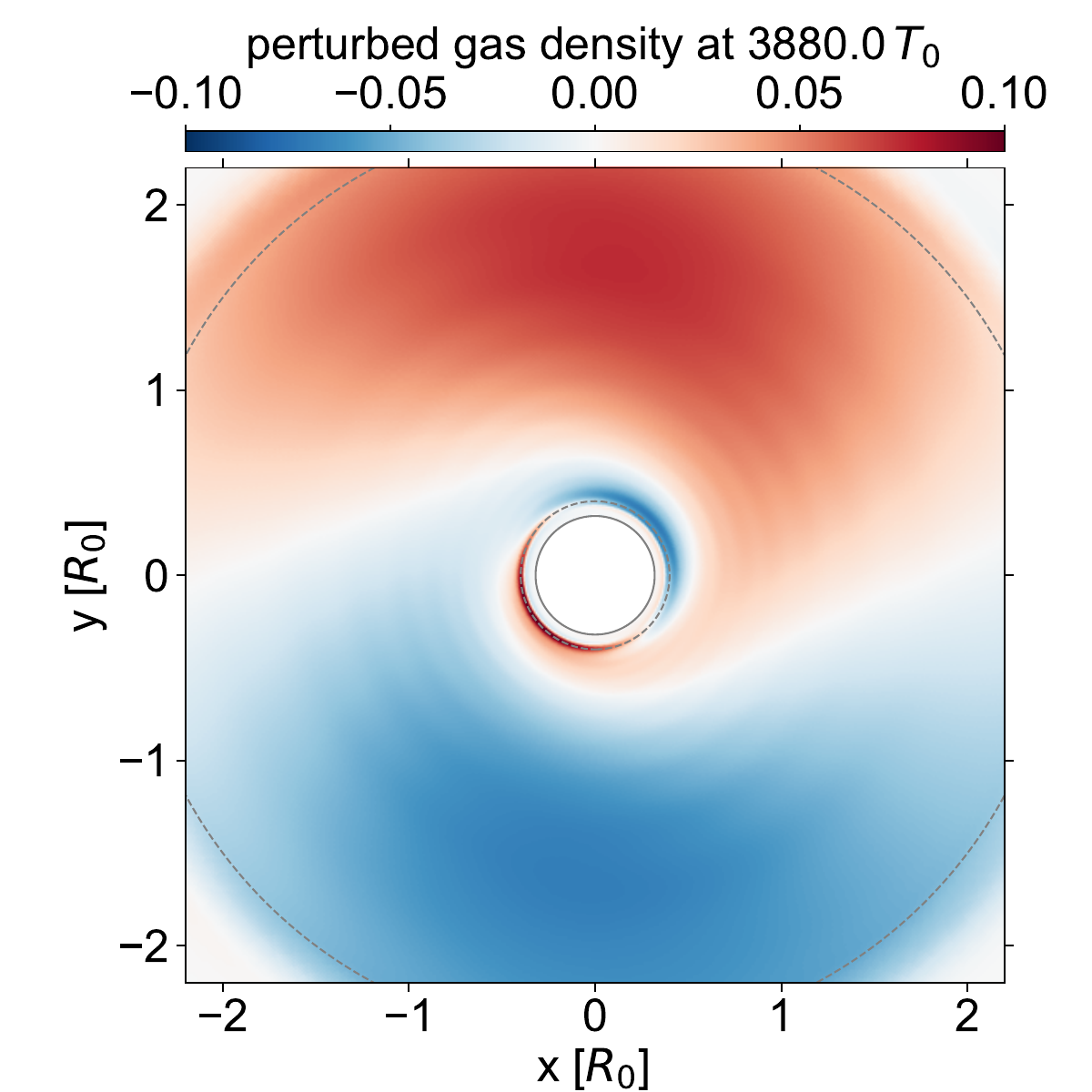}
    \caption{Results of our reference setup for the \fargotwodoned\ simulation which uses a barycentric frame. The left panel shows the position of the star in the barycentric frame. The open square symbols mark the star's position in the middle and right panels, and the arrows display the acceleration exerted by the disc on the star (arbitrary magnitude). The middle and right panels display the perturbation of the surface density of the disc relative to its azimuthally-averaged initial density, at two times separated by about half a pattern period. The solid curve shows the inner edge of the computational grid, the dashed curves show the edges of the inner and outer wave-damping zones.}
    \label{fig:bary}
\end{figure*}
%FFFFFFFFF

\begin{itemize}
\item[\sbt] \textcolor{blue}{\fargotwodoned}\ \citep{Crida+2007}, a sibling of the original \fargo\ code \citep{FARGO} which solves the governing equations in the barycentric frame. We did not make use of the 1D grid in the simulations presented here.
\item[\sbt] \href{https://github.com/charango/dustyfargoadsg}{\fargoadsg\ } \citep{BaruteauMasset2008AD, BaruteauMasset2008SG}, an extension of \fargo\ used to study the impact of self-gravity on the growth of the instability. Contrary to \fargotwodoned, \fargoadsg\ solves the governing equations in the stellocentric frame. 
\item[\sbt] \href{https://fargo3d.bitbucket.io/}{\fargothreed\ } \citep{BenitezLlambay2016fargo3d}, the 3D successor of \fargo, which is used to examine the instability in 3D (the code also employs a stellocentric frame).
\item[\sbt] \href{https://github.com/idefix-code/idefix}{\idefix\ } \citep{Lesur2023Idefix}, which, contrary to the \fargo\ family codes, 
is a code based on a higher order Godunov scheme that uses a second-order Runge-Kutta time integrator to solve the governing equations in the stellocentric frame.
\item[\sbt] \href{https://phantomsph.github.io/}{\phantomcode\ } \citep{Price2018Phantom}, a 3D SPH (Smoothed Particle Hydrodynamics) code which, like \fargotwodoned, employs a barycentric frame but, unlike all previous grid-based codes, has the obvious advantage of having no boundaries. However, SPH particles moving inside an accretion radius centred on the star are removed and their mass and momentum are transferred to the sink particle representing the star. The discrete initial position of the SPH particles plays the equivalent role of white noise in our grid-based simulations to break the disc's axisymmetry (though not at the same 0.1\% level).
\end{itemize}
By default, in the grid-based simulations carried out in a stellocentric frame, the disc's indirect term \ITdd, which is the opposite of the total acceleration exerted by the disc on the star (see Paper~I), is applied to each grid cell.
The 2D grid-based simulations use 460 grid cells with a logarithmic spacing along the radial direction and 1256 cells along the azimuthal direction, ensuring a resolution $dR/R=d\varphi=0.005$. 
The 3D run performed with \fargothreed\ has 230$\times$628$\times$48 cells in the radial, azimuthal and latitudinal directions, respectively, with latitudes spanning $\pm0.15$ rad about the disc midplane.
So-called wave-damping zones are used for $R \in [0.32-0.4]R_0 \cup [2.5-3.125]R_0$, where the disc fields are damped towards their initial axisymmetric profile.
The \phantomcode\ simulation uses $5\times10^5$ particles, with the sink accretion radius taken equal to $r_\mathrm{sink}=0.1R_0$.
In \fargoadsg simulations with gas self-gravity, the self-gravitating acceleration uses a smoothing length\footnote{Since the disc is isothermal, the smoothing length in the self-gravitating acceleration does not scale linearly with $R$, which is yet required for the self-gravitating acceleration to be formally expressed as a convolution product and be computed with fast Fourier transforms \citep{BaruteauMasset2008SG}. Using a smaller smoothing length, or a different disc model with uniform Toomre-Q parameter and uniform aspect ratio, does not qualitatively change the results of self-gravitating simulations presented in \S\ref{sec:interpretation}.} of $0.3H(R)$, with $H(R) = h(R) \times R$ the disc's pressure scale height.

% ================
\section{Overview of the instability}
\label{sec:overview}
% ================

\begin{table}[]
    \centering
    \begin{tabular}{|c|c|c|c|c|}
      \hline
      Code & Frame & 2D/3D & Self-gravity & $\tau_\mathrm{g}$ ($T_0$) \\ \hline\hline
      \fargotwodoned & barycentric & 2D & No & 300 \\ \hline
      \phantomcode & barycentric & 3D & No & 62  \\ \hline
      \multirow{2}{*}{\fargothreed} & \multirow{2}{*}{stellocentric} & 2D & \multirow{2}{*}{No} & 83 \\ 
      && 3D && 130 \\ \hline
      \idefix & stellocentric & 2D & No & 118 \\ \hline
      \multirow{2}{*}{\fargoadsg} & \multirow{2}{*}{stellocentric} & \multirow{2}{*}{2D} & No & 91 \\ 
      &&& Yes & 27 \\ \hline
    \end{tabular}
    \caption{Differences between the codes and numerical set-ups used in
    our simulations.
    Column~5 lists the associated exponential growth timescale $\tau_\mathrm{g}$.}
    \label{tab:code_params}
\end{table}

% ----------------------------
\subsection{Results of the reference setup}
\label{sub:reference_setup}
% ----------------------------

We present in \figref{fig:bary} the results of a simulation for our reference setup with \fargotwodoned, which employs a barycentric frame.
The left panel shows that the star undergoes an outward spiraling motion around the centre-of-mass of the system. 
This outward spiraling motion is associated with the growth of a global mode with azimuthal wavenumber $m=1$, as can be seen in the perturbed density in the middle and right panels.
Between the middle and right panels, which are separated by 100 $T_0$, the global mode has rotated by about $180^\circ$, and it is clear from the left panel that this corresponds to about half the period of the star's spiraling motion.

We point out that the growing perturbed density varies in phase with orbital radius; for instance, in the right panel, the bulk of the disc has a negative (positive) density perturbation in the southern (northern) half of the disc, but it is not the case close to the grid's inner edge.
The disc's outer regions set the position of the centre-of-mass, so that the star is located opposite to the large, smooth positive perturbed density at both times. However, the disc's inner regions set the acceleration exerted by the disc on the star, as shown by the blue arrows in the left panel of the figure, which both point toward the positive perturbed density near the disc's inner edge.

%FFFFFFFFF
\begin{figure*}
    \centering
    \includegraphics[width=\columnwidth]{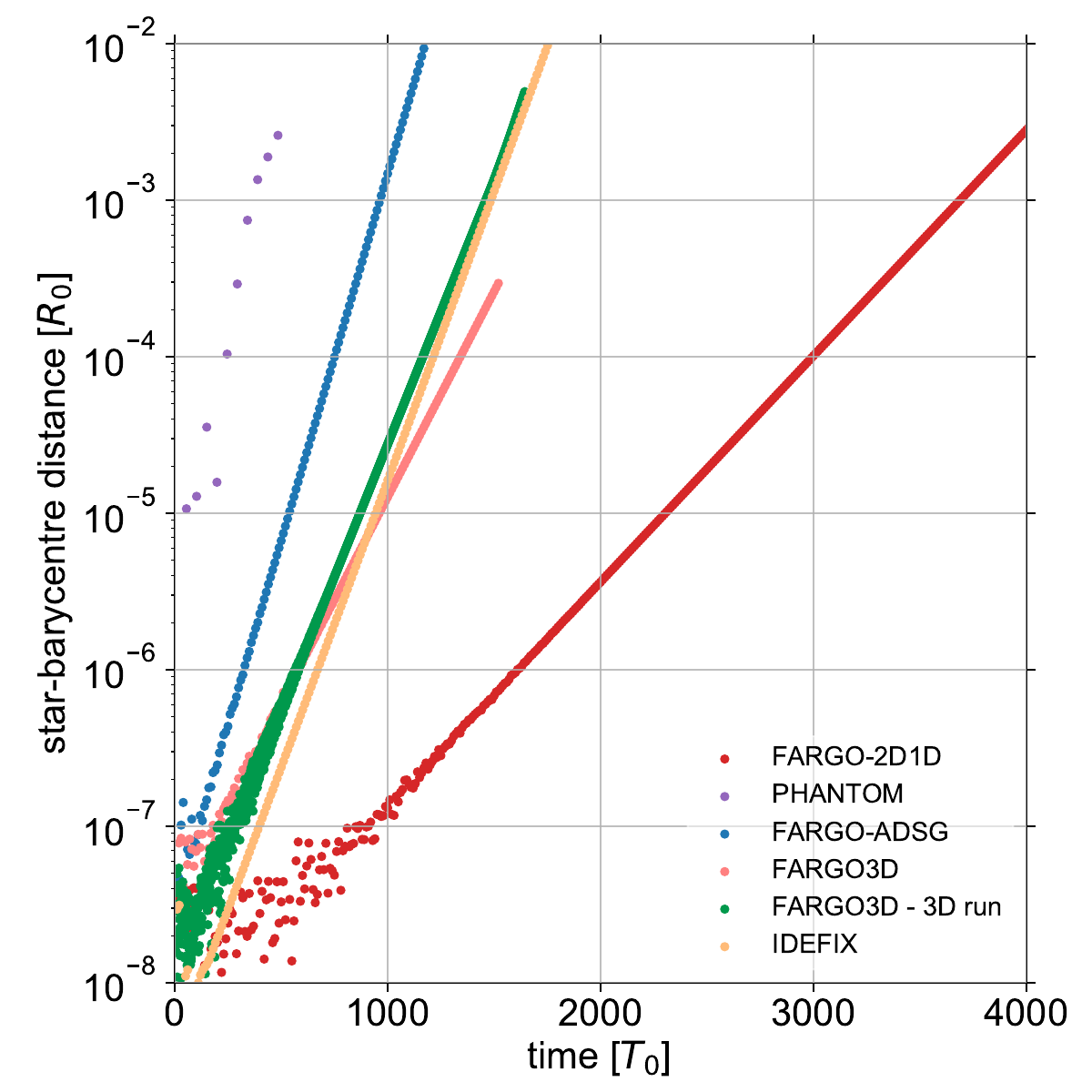}
    \includegraphics[width=\columnwidth]{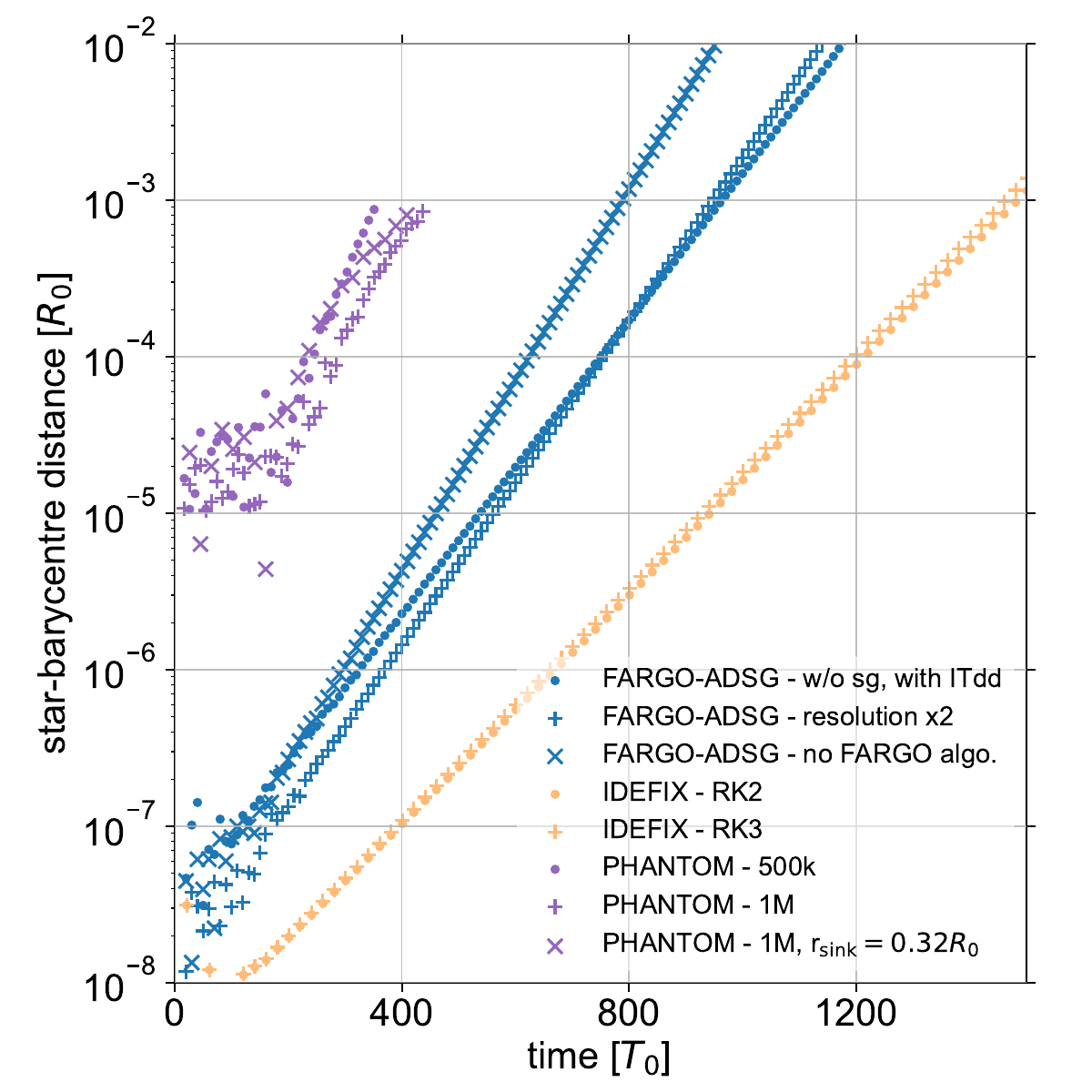}
    \caption{Time evolution of the radial distance between the star and the centre-of-mass of the star-disc system for simulations using our reference setup, obtained for different codes and by varying numerical parameters (see caption and text).}
    \label{fig:com_tlogr_numerics}
\end{figure*}
%FFFFFFFFF

To illustrate this spiraling motion, \figref{fig:com_tlogr_numerics} shows the distance between the star and the centre-of-mass of the star and disc system as a function of time. The red dots on the right of the left panel show that it grows exponentially with a timescale $\tau_\mathrm{g}\approx300\ T_0$.

% ================
\subsection{Numerical robustness}
\label{sub:numeric}
% ================

%\paragraph{Solver, algorithm and frame}

To confirm this puzzling result, we carried out simulations using our reference set-up with other codes. Table~\ref{tab:code_params} lists their differences. All simulations presented in this section neglect the disc self-gravity. The simulation run with \phantomcode\ in 3D yields qualitatively the same result as shown in the previous subsection with \fargotwodoned, with the exponential growth of an $m=1$ mode in the disc, though the growth timescale is reduced to $\tau_\mathrm{g}\approx 60\ T_0$ (see the left panel of \figref{fig:com_tlogr_numerics}, purple dots on the left).
The other codes, \fargoadsg, \idefix, \fargothreed\ (in 2D and in 3D), which use a stellocentric frame, all find the exponential growth of this distance too. It is shown in the left panel of \figref{fig:com_tlogr_numerics}, and is equivalent to the amplitude of the $m=1$ mode in the disc.
This effect is therefore independent of the reference frame, and appears in simulations using codes of various types and solvers.

%\paragraph{Parameters}

To assess whether this phenomenon could be due to a numerical effect common to all the codes used above (like a bad implementation of the Indirect Term, or the boundary conditions), we ran additional simulations with the same physical set-up, by changing a few numerical parameters. The right panel in \figref{fig:com_tlogr_numerics} illustrates the results obtained when increasing resolution in \phantomcode\ and \fargoadsg, or removing the FARGO algorithm (hence shortening the hydrodynamical time-step) in \fargoadsg, or modifying the numerical scheme of integration in \idefix. The results are consistently independent of these numerical parameters.

We performed further tests to assess the impact of the boundary conditions for the grid codes. For instance, we ran simulations where the disc's axisymmetry was broken by the inclusion of a planet on a fixed orbit instead of white noise, and using a non-reflecting boundary condition that avoids reflections of the planet's wakes off the grid's edges instead of wave-damping zones. We also ran simulations where the disc's initial density was smoothly dropped by a large factor ($\approx 100$) near the edges of the grid. Finally, we ran simulations with \fargotwodoned\ that used an additional 1D grid surrounding the 2D grid along with the use of open 'outflow' boundary conditions at the edges of the 1D grid (David Fang, private communication). In all cases, we always found an exponential growth of the star-barycentre distance. We further stress that \phantomcode\ has no boundaries stricto-sensu, but the sink particle would be equivalent to an open boundary condition at a radius ($0.1R_0$ in the default case) smaller than the inner edge of the grid in our simulations.
Besides, we note that real protoplanetary disks are limited in space, and their physical boundaries may be regions of wave reflection similar to possible wave reflection at the edge of the grid in the simulations.

Finally, we varied several disc parameters and found that the growth timescale is a non-trivial combination of the disc's surface density and aspect ratio at the inner edge. 
The use of wave-damping zones and their detailed implementation or, in the specific case of \fargotwodoned, the fact that the inner edge of the grid is centred on the centre of mass of the system and not on the star, may affect the physical conditions at the effective inner edge radius.
A last \phantomcode\ simulation, using $r_\mathrm{sink}=0.32R_0$ and plotted in the right panel of Fig.~\ref{fig:com_tlogr_numerics}, shows a longer $\tau_\mathrm{g}\sim110T_0$ more in line with the grid simulations, which would support this interpretation.
We think that this explains the quantitative differences observed in our simulations (in the results shown in \figref{fig:com_tlogr_numerics}, 
the growth timescales are $\tau_\mathrm{g}=113 T_0 \pm 0.2$ dex). In particular, we note that the two codes with the most extreme values of the growth timescale are \phantomcode\ and \fargotwodoned, that is the ones with a different geometry of the inner edge.
In the next section, we present arguments in favor of a physical origin for the qualitatively consistent appearance of the exponential growth of the star-barycentre distance in all our simulations.

%FFFFFFFFF
\begin{figure*}[ht]
    \centering
    \includegraphics[width=0.33\hsize]{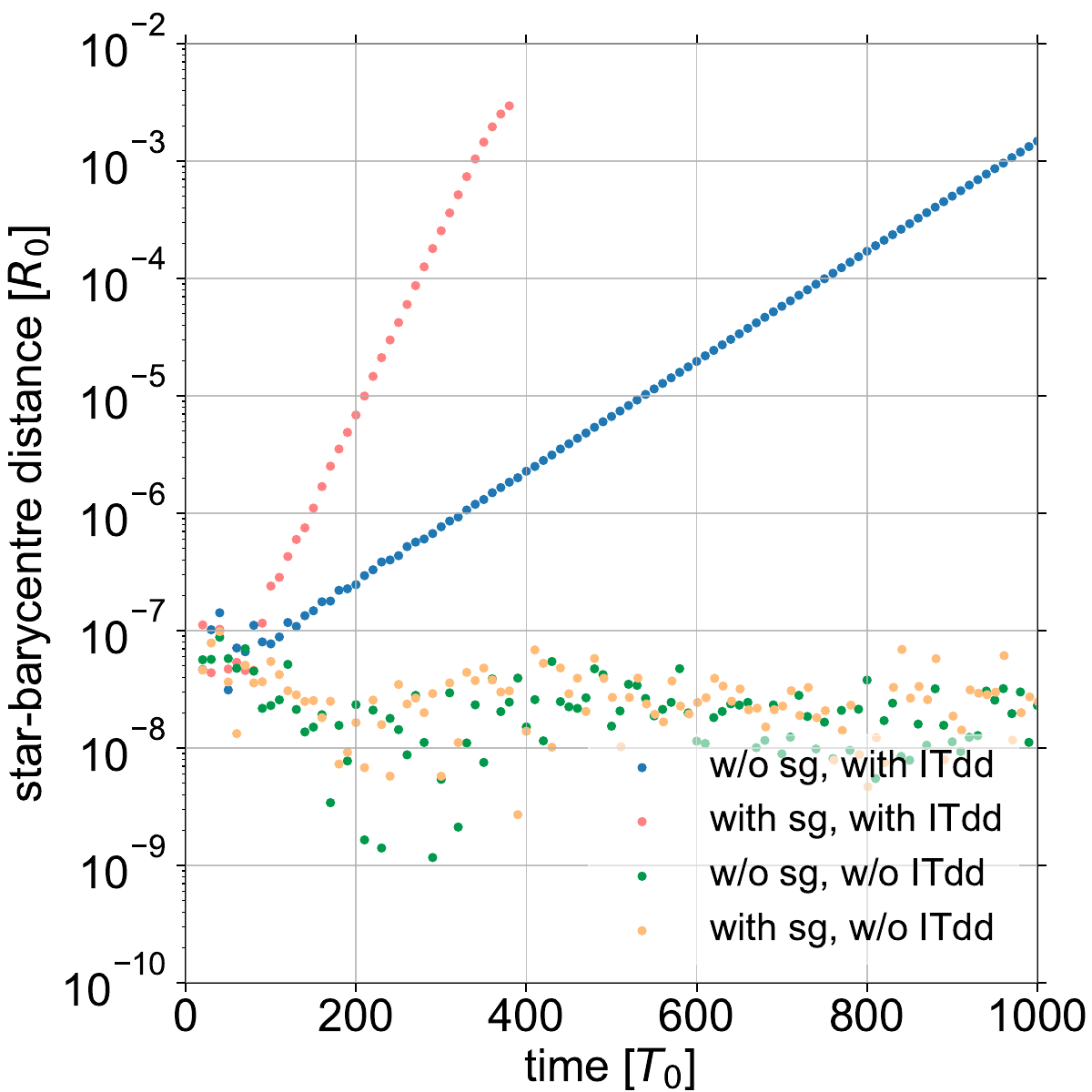}
    \includegraphics[width=0.33\hsize]{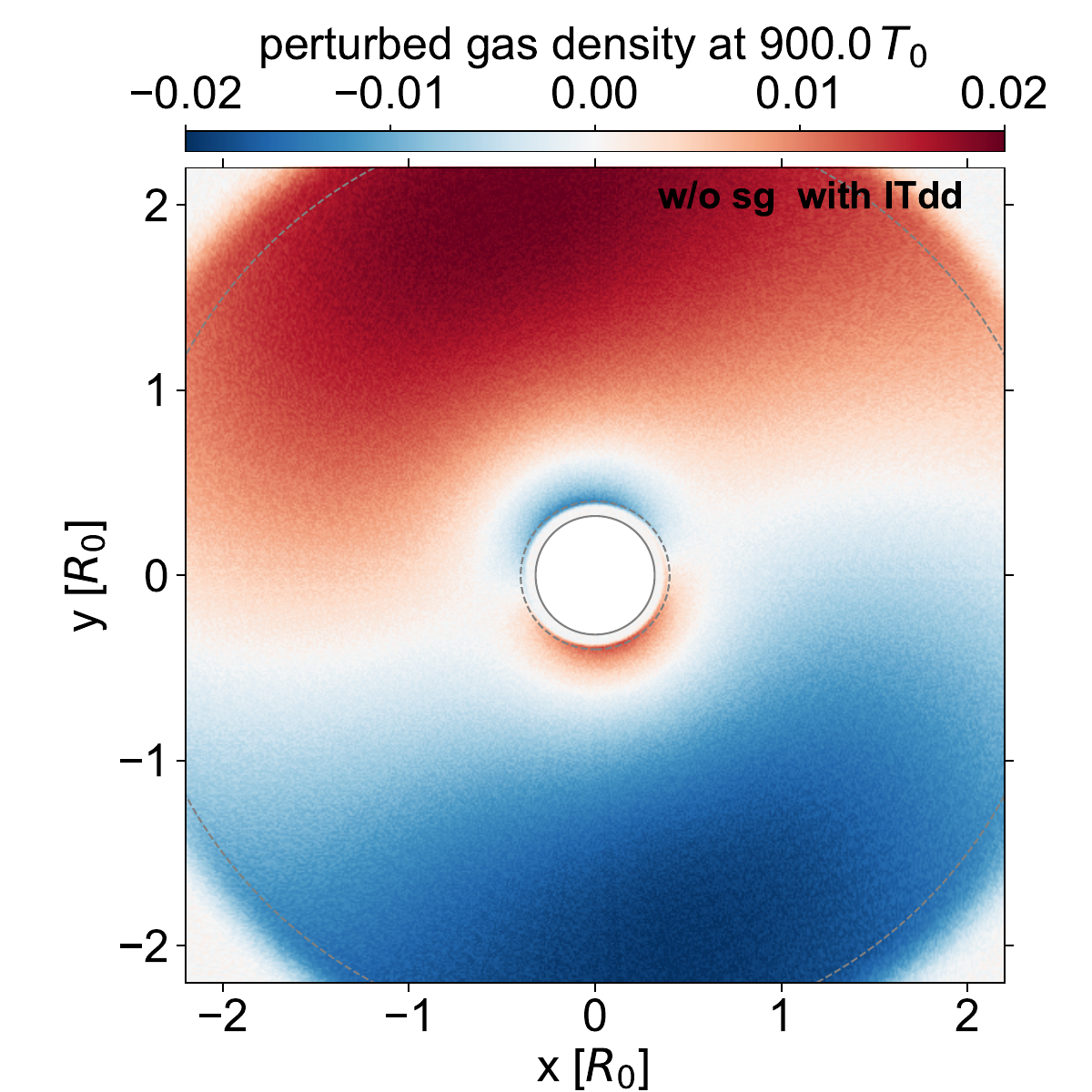}
    \includegraphics[width=0.33\hsize]{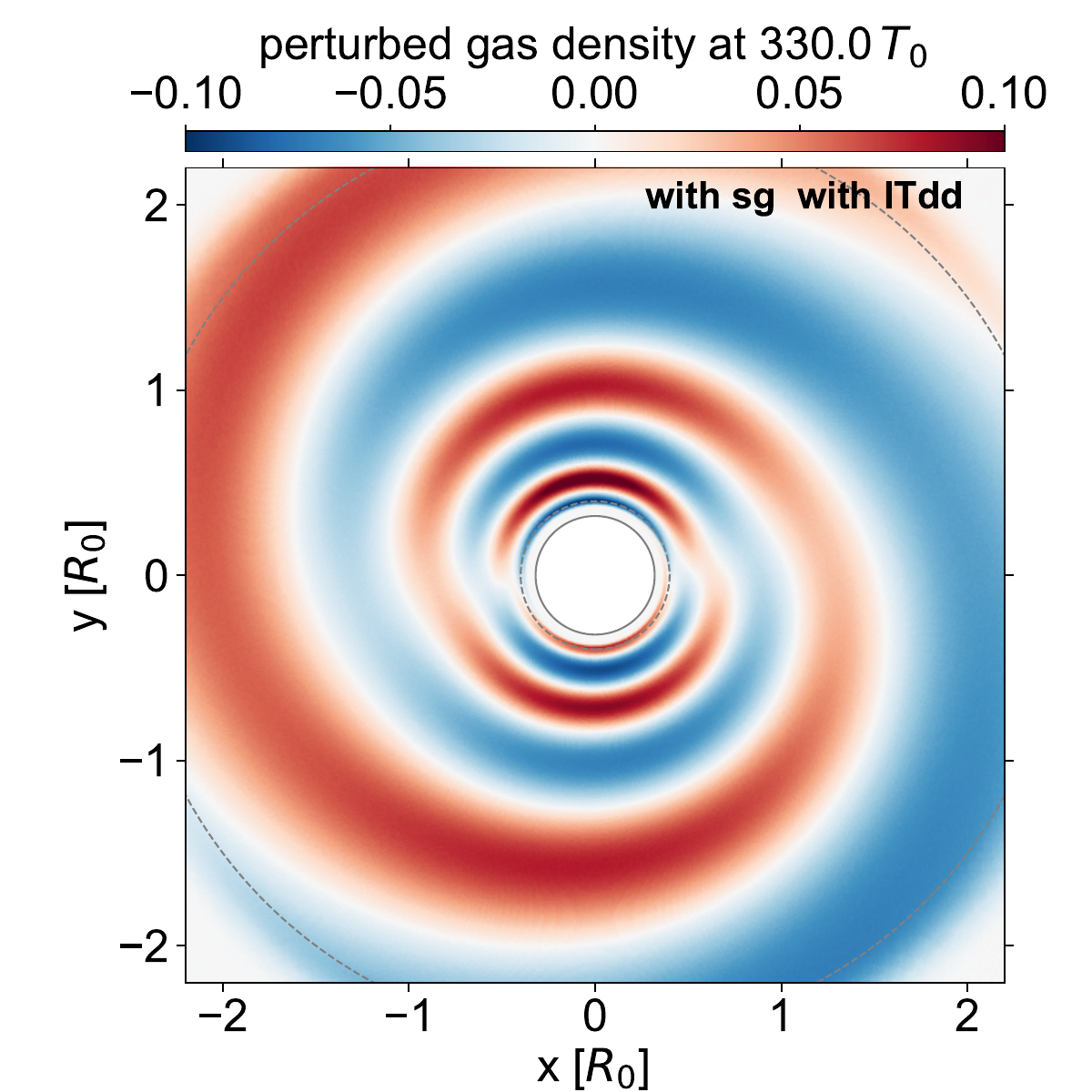}
    \caption{Results of our reference setup for \fargoadsg\ simulations using a stellocentric frame. The left panel shows the time evolution of the radial distance between the star and the centre-of-mass of the star-disc system for four simulations with or without self-gravity ('sg'), and with or without the indirect term of the disc on itself ('ITdd'). The middle and right panels display the perturbation of the surface density of the disc relative to its azimuthally-averaged initial density for the simulations with \ITdd, without self-gravity (middle panel) and with self-gravity (right panel). The solid curve shows the inner edge of the computational grid, the dashed curves show the edges of the inner and outer wave-damping zones.\vspace{10pt}}
    \label{fig:sg}
\end{figure*}
%FFFFFFFFF

% ----------------------------
%\subsection{Preliminary characterisation of the instability}
% ================
\section{Physical interpretation}
\label{sec:interpretation}
% ================
\paragraph{A positive feedback loop}
Previous sections strongly suggests that the instability found using either a stellocentric frame with \ITdd\ or a barycentric frame is of physical origin.
However, when the same simulations are run in the stellocentric frame without \ITdd, the disc remains stable for ever. It is expected indeed for this set-up that the disc is intrinsically stable. Similarly, a barycentric simulation in which the star does not feel the disc acceleration remains stable for ever. This shows that the instability originates in all the simulations from a positive feedback loop between an asymmetry in the disc and the reflex motion of the star.

Further insight into the feedback loop can be gained by integrating numerically the motion of a test particle that is initially on a circular orbit around a star. 
When applying a constant acceleration on the star, the eccentricity of the test particles rises linearly, with the longitude of pericentre shifted by $-\pi/2$ with respect to the acceleration vector \citep[see also][]{Namouni-2005}. 
Now, if the acceleration is set proportional to the particle's eccentricity, and phase-shifted by $+\pi/2$ from the direction of pericenter, we find an exponential increase in the particle's eccentricity (see Appendix A). Similarly, in hydrodynamical simulations where we apply a constant and uniform acceleration to the whole disc (like the indirect term), the disc eccentricity grows and an $m=1$ mode appears in the perturbed density. In turn, this $m=1$ mode exerts a force on the star\footnote{$m>1$ modes result in zero acceleration of the star.} which is proportional to the amplitude of the mode (thus to the eccentricity) and, as pointed out in \S~\ref{sub:reference_setup}, may not point towards the centre of mass of the system if this mode has a phase that varies with radius. 
If the resulting acceleration has a component pointing at $+\pi/2$ from the longitude of the pericentre at a given radius, the loop is closed.

\paragraph{Role of self-gravity}
All the simulations presented in section~\ref{sec:overview} suffer however --\,like most disc simulations in the literature\,-- from an inconsistency which we put forward in Paper~I\,: while the star feels the gravitational acceleration from the disc, the disc does not, because we do not include its self-gravity. This imbalance can be problematic, and typically makes tidal effects impossible to account for. Therefore, we have run our reference set-up with \fargoadsg, with the gas self-gravity on. The results are shown in \figref{fig:sg}, right panel, compared to the middle panel where the self-gravity is off, at different times corresponding to a similar amplitude of the mode. 
In both cases, the star-barycentre distance grows exponentially, as shown in the left panel, and only the timescale and the shape of the mode change. The growth timescale $\tau_\mathrm{g}\approx27\;T_0$ for the self-gravitating run, that is $\approx3.3$ times shorter than for the same run without self-gravity. Conversely, in a simulation with only self-gravity but no \ITdd, the disc remains stable. Of course with neither self-gravity nor \ITdd, nothing happens either.
Therefore, the feedback loop described above is \emph{not} caused by the absence of self-gravity in the disc. The stellar acceleration due to the disc is sufficient and necessary to trigger this instability. Self-gravity then modifies the response of the disc quantitatively.

\begin{figure}
    \centering
    \includegraphics[width=0.8\hsize]{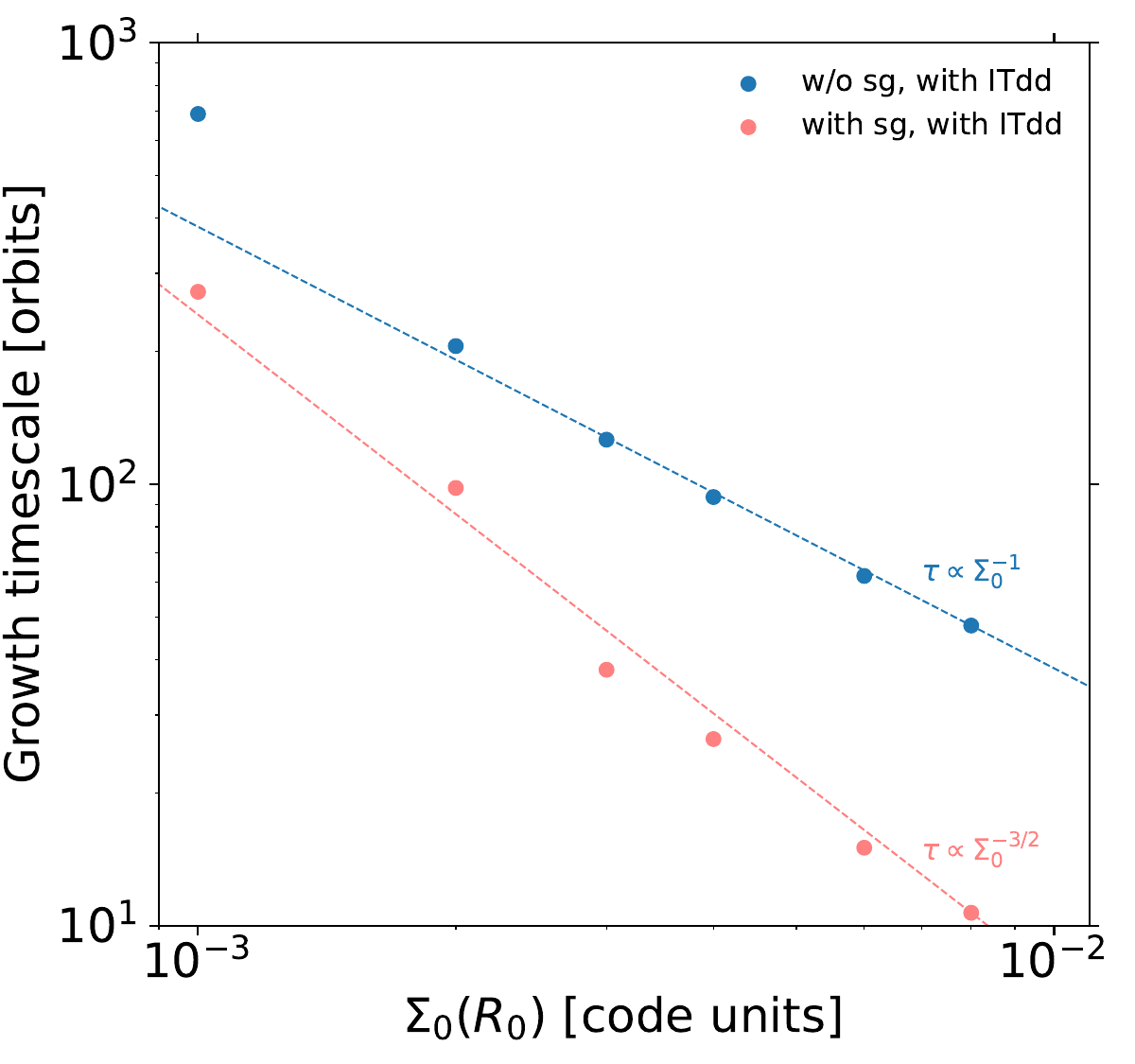}
    \caption{Growth timescale $\tau_g$ versus the initial surface density $\Sigma_0$ at the reference radius $R_0$, with and without self-gravity. The dashed lines show suggestive trends.}
    \label{fig:varys0}
\end{figure}

\paragraph{Dependence with the disc's surface density}
While a detailed characterisation of the instability is beyond the scope of this Letter and is deferred to a future, on going study, we briefly present the results of simulations where we varied the initial surface density at the reference radius ($\Sigma_0$).
Such simulations were carried out with \fargoadsg, with and without self-gravity, and the uniform Toomre-Q parameter varies from 2 to 16 across simulations.
For each simulation, the growth timescale $\tau$ is inferred from the time evolution of the star-barycentre distance. 
\Figref{fig:varys0} shows that $\tau\propto 1/\Sigma_0$ for non-self-gravitating runs, as expected for a linear perturbation where the feedback is proportional to $\Sigma_0$.
Self-gravitating simulations show a slightly steeper dependence with $\Sigma_0$ ($\tau\propto\Sigma_0^{-3/2}$).\vspace{-8pt}

%FFFFFFFFF
%FFFFFFFFF

% ================
\section{Discussion}
\label{sec:discussion}
% ================

We have found that non-axisymmetric Keplerian discs are linearly unstable against the growth of an $m=1$ mode arising from the reflex motion of the star around the centre-of-mass of the star and disc system. 
For this reason we shall dub it {\it reflex instability} for future reference.

We identified a feedback loop which could be the cause of this instability, and described how the disc eccentricity and the acceleration of the star can excite each other. However, due to the lack of a fundamental linear analysis so far (work in progress), there is still a risk that this instability is of numerical origin. Even if it were the case, we stress that it seems to be ubiquitous in numerical codes, and we warn the community that they may encounter this exponential growth of an $m=1$ mode in the disc in their long-term simulations that include \ITdd.

At first glance it may seem surprising that the reflex instability has not been reported before for typical protoplanetary disc masses. 
The explanation could be two-fold.
On the one hand, for typical disc-to-star mass ratios below a few percent, the growth timescale can exceed $10^3$ dynamical timescales, thereby requiring rather long-term simulations in order not to fly under the radar.
Actually, some of us (and other colleagues, e.g. G. Pichierri who observed the growth of an oscillation of the torque felt by a planet around its average value, private communication) have been surprised in the past by the sudden and unexplained excitation of the disc and/or crash of the code in very long term simulations; in retrospect, we think this strange behaviour can be attributed to the reflex instability, whose amplitude was negligible for most of the time. 
On the other hand, most simulations of star-disc interactions use a stellocentric frame and thus require the disc's indirect term to be activated for the instability to manifest itself, and it turns out that the inclusion of indirect terms has long remained nebulous (see Paper~I). In fact, in most non-self-gravitating simulations of discs without planets, \ITdd\ is not included \citep[e.g.][]{Robert+2020}.

This reflex instability shares characteristics with the work of \citet{Heemskerk92}. They did a linear stability analysis of $m=1$ disturbances showing that the shift of the star away from the system's centre-of-mass leads to the growth of an instability in large ($\gtrsim 1$) disc-to-star mass ratios. This instability was further confirmed by non-linear hydrodynamical simulations for similarly large disc-to-star mass ratios. The authors concluded that the instability disappears when the disc mass becomes lower than the star mass.
We note however that their linear analysis actually discards collective effects (namely self-gravity and pressure), so that it does not apply to our framework. In their section 3.5, the authors state that if self-gravity and pressure were included in the disc response, an instability may be allowed to persist ``at some level'' for any disc-to-star mass ratio. Furthermore, in their numerical simulations, they report finite growth rates even for their lowest disc-to-star mass ratio, $1/9$, which is barely twice the value we use in our simulations.
A dedicated linear analysis in the regime of low disc-to-star mass ratios will be presented in a future study.

Interestingly, \citet{Peng-Batygin-2020} have observed a growth of the eccentricity in pure N-body simulations containing only \emph{small particles}, that is particles that have a small mass but do not feel each other's direct gravity. As stated by the authors, this unexpected dynamical excitation stems from the reflex motion of the star due to the global motion of the cloud of small particles.
However, since the small particles do not have a collective behaviour, no $m=1$ mode can grow so that in their case, there is no feedback loop and no exponential growth.

Also, \citet{Saha+2007} have studied the persistence of a $m=1$ mode in galactic discs with an exponentially decreasing surface density profile and no central mass, but with self-gravity. Although the physical set-up is different from ours and the rotation velocity is far from Keplerian, they find that the mode precesses very slowly and that "\emph{the disc is never unstable to the lopsided $[m=1]$ modes without the indirect term arising due to the lopsided perturbation itself}". The reflex instability therefore seems to have astrophysical applications well beyond the field of protoplanetary discs and to be a quite generic phenomenon.

In this paper we have only explored the linear growth stage of the reflex instability, intentionally leaving its non-linear development and its full characterisation for future work.
Note that the grid-based simulations of the present study did not show any saturation of the instability, until the disc reaches quite extreme eccentricities, causing most often a crash of the simulation.
The growth of a global $m=1$ mode could limit the lifetime of massive discs and excite the disc's eccentricity with interesting implications for scenarios of planet formation and orbital evolution. 
It may also imply the formation of a large-scale spiral throughout the disc, with potential connections to spiral structures observed in the gas emission of protoplanetary discs.

%=======================%
\section*{Acknowledgments}
%=======================%
We thank very much several colleagues for enlightening discussions on this subject: William Béthune, Alessandro Morbidelli, Gabriele Pichierri, Nathan Magnan, Fabiola Gerosa, Dylan Kloster, Kate Minker, Léa Marques, Thomas Rometsch and Lucas Jordan.
We also thank the anonymous referee whose comments helped to improve this manuscript.

We would like to warmly thank the "Programme National de Physique Stellaire" (PNPS) and "Programme National de Planétologie" (PNP) of CNRS/INSU co-funded by CEA and CNES for their financial support of our BiPhasique research group. JFG thanks the LABEX Lyon Institute of Origins (ANR-10-LABX-0066) for its financial support within the Plan France 2030 of the French government operated by the ANR. FM acknowledges support from UNAM's grant PAPIIT 107723, UNAM's DGAPA PASPA program and the Laboratoire Lagrange at Observatoire de la C\^ote d'Azur for hospitality during a one-year sabbatical stay. Part of the numerical simulations were performed on the CALMIP Supercomputing Centre of the University of Toulouse, and others on “Mesocentre SIGAMM” hosted by Observatoire de la Côte d’Azur.
The authors are also grateful to the Université Côte d’Azur’s Center for High-Performance Computing (OPAL infrastructure) for providing resources and support.
We acknowledge support by DFG-ANR supported GEPARD project (ANR-18-CE92-0044 DFG: KL 650/31-1).
This work was supported by the French government through the France 2030 investment plan managed by the National Research Agency (ANR), as part of the Initiative of Excellence Université Côte d’Azur under reference number ANR-15-IDEX-01.
%=======================%

\begin{appendix}

\section{Link between acceleration and eccentricity}

\subsection{Laplace vector, eccentricity, and acceleration}
\label{sub:Laplace}

For a particle in orbit around a central object of mass $M$, we remind that the Laplace vector is defined as 
$\displaystyle \vec{P}=\frac{1}{GM}(\vec{v}\times\vec{h})-\vec{u}_r$,
where $\vec{v}$ is the velocity of the test particle, $\vec{h}=\vec{r}\times\vec{v}$ is its specific orbital angular momentum, and $\vec{u}_r$ is the unit vector in the radial direction.
One can show that $\vec{P}$ is constant\,: its norm is equal to the eccentricity of the orbit and its direction points towards the pericentre. Without loss of generality, we choose here the longitude of the pericentre $\varpi$ to be zero, so that $\vec{P}=e\,\vec{u}_x$, where $\vec{u}_x$ is the unit vector in the $x$-direction ; $\vec{u}_y$ and $\vec{u}_z$ complete an orthonormal base with $\vec{u}_z$ perpendicular to the orbital plane, such that $\vec{h}=\sqrt{GMa(1-e^2)}\vec{u}_z$ (where $a$ is the semi-major axis).

From the definition of $\vec{P}$ above, one can show that
\begin{equation}
    \vec{v} =  \sqrt{\frac{GM}{a(1-e^2)}}\left(\vec{u}_\phi + e\vec{u}_y \right) = \sqrt{\frac{GM}{a}}\big(\vec{u}_\phi + e\vec{u}_y + \mathcal{O}(e^2)\big)\ .
    \label{eq:v}
\end{equation}
In other words, for small eccentricities, the velocity is decomposed into a uniform circular motion ($\vec{u}_\phi$ is the unit vector in the azimuthal direction) and a uniform linear motion parallel to the minor axis of the ellipse.\vspace{12pt}

Hence, applying a constant acceleration $\vec{a}_c=\varepsilon\vec{u}_y$ (e.g. an indirect term) to a particle on an initially circular orbit should lead to an increase of the linear component of the velocity $v_y=\sqrt{\frac{GM}{a}}e$, that is an increase in the eccentricity as $\dot{e}=\varepsilon/\sqrt{GM/a}$.

\begin{figure}
  \begin{center}
    \includegraphics[width=0.48\linewidth]{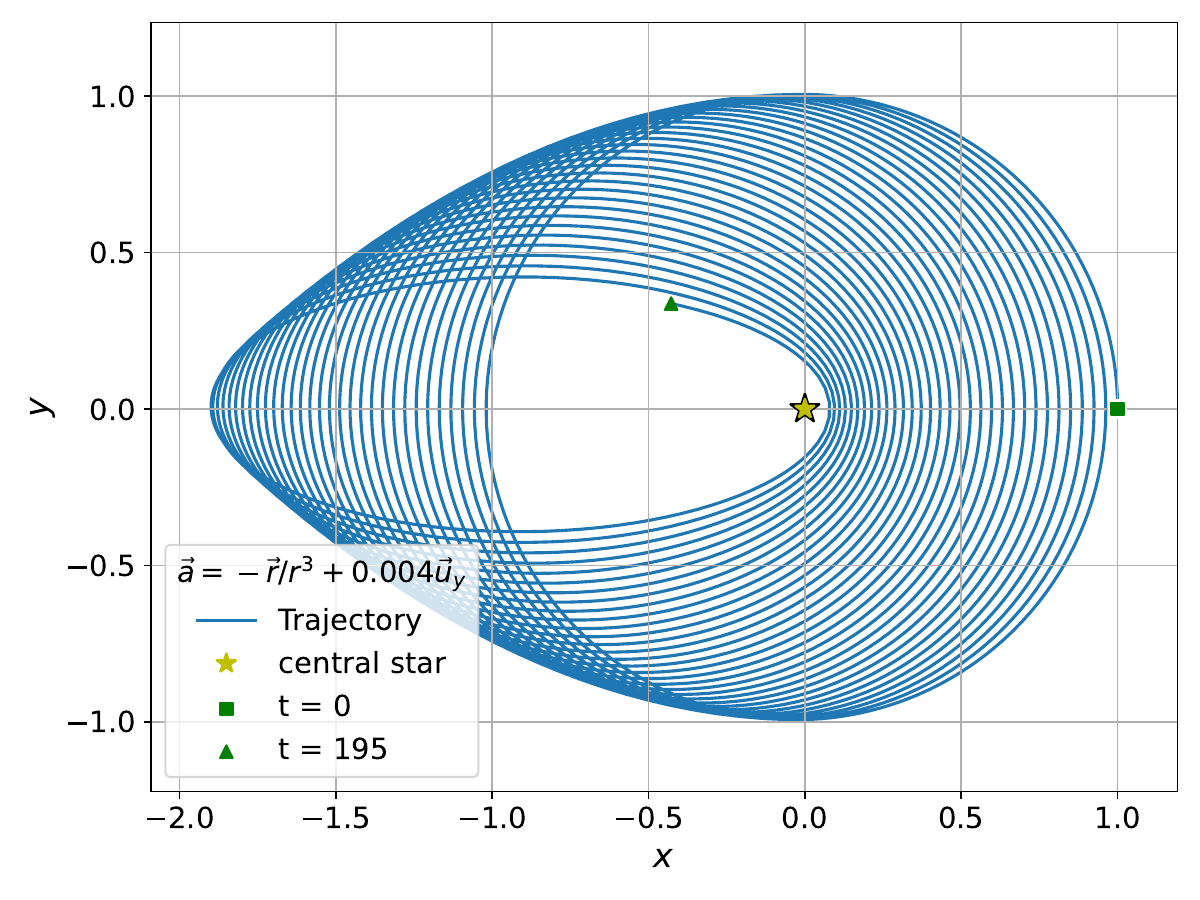}
    \includegraphics[width=0.48\linewidth]{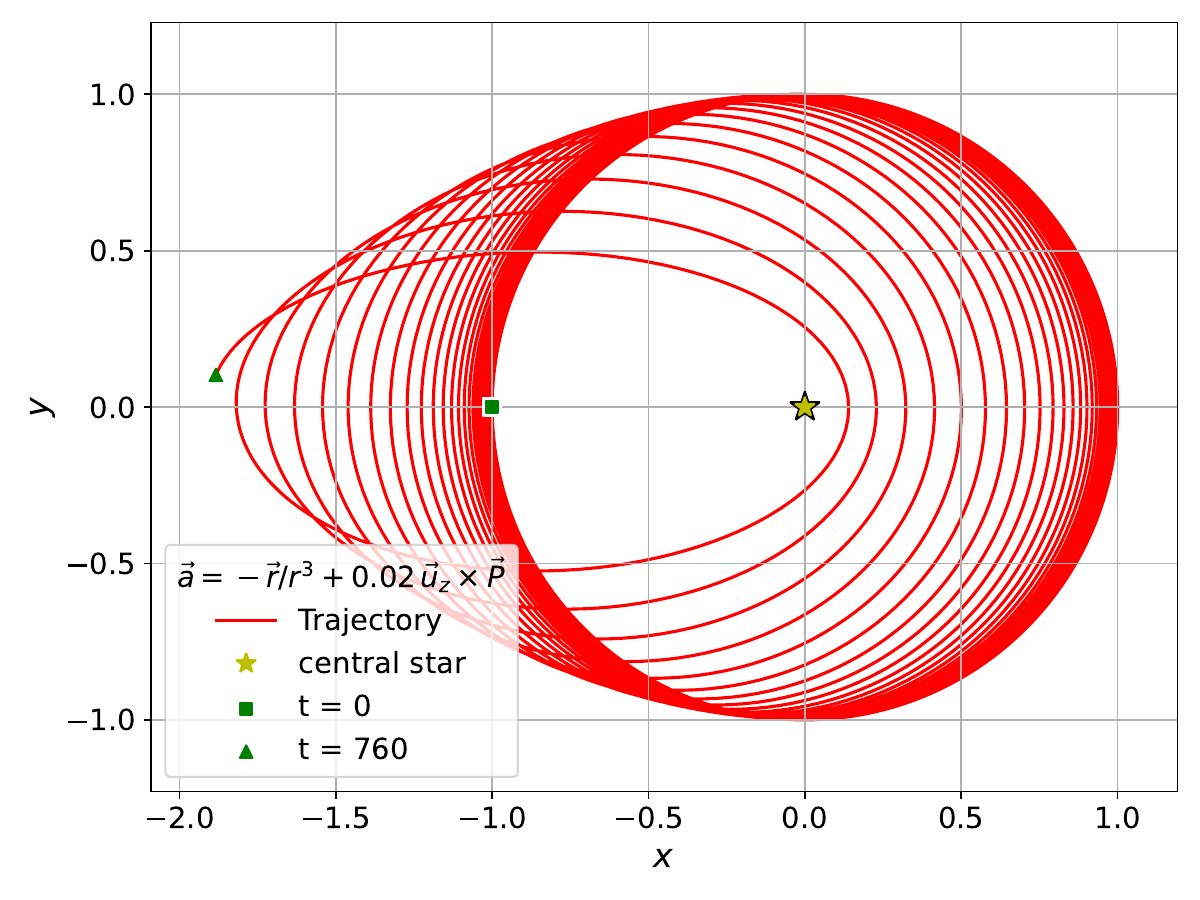}\\
    \includegraphics[width=0.48\linewidth]{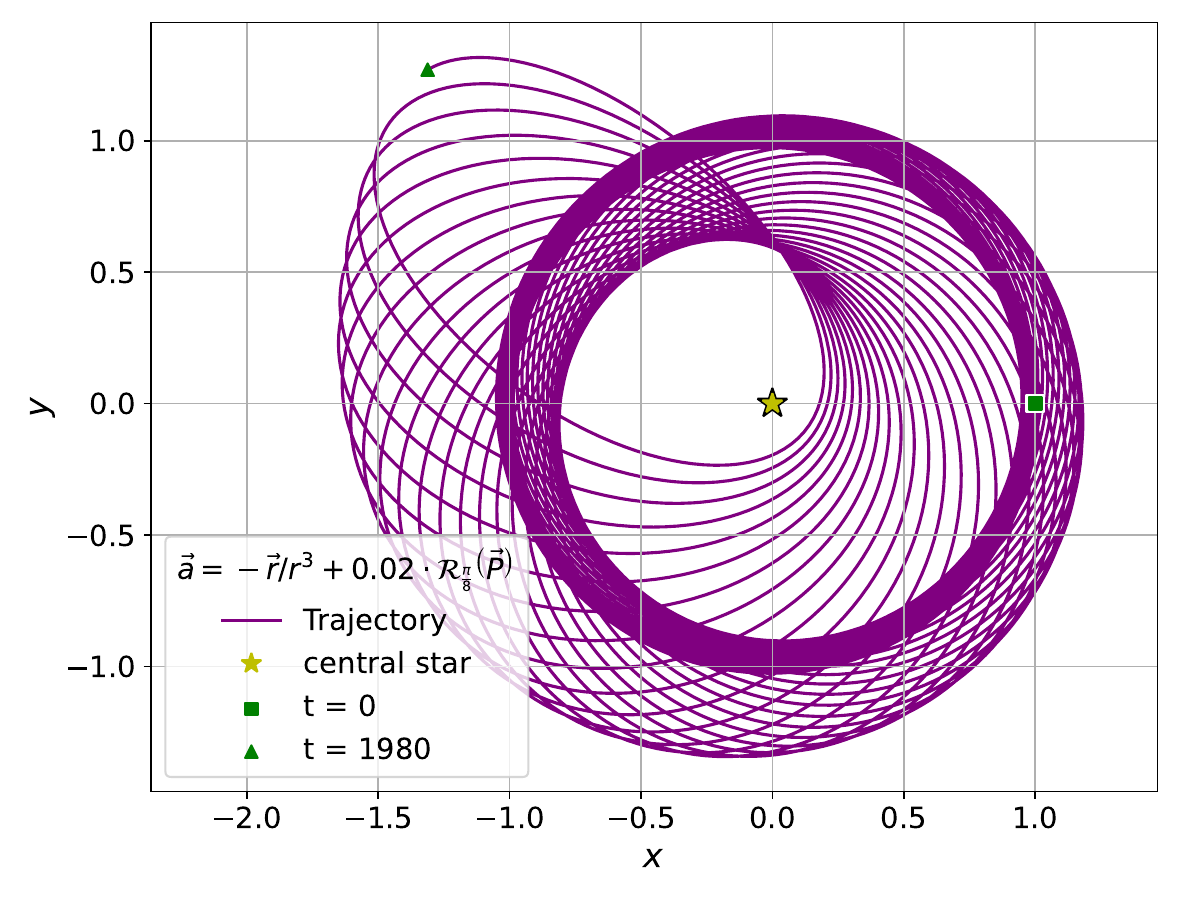}
    \includegraphics[width=0.48\linewidth]{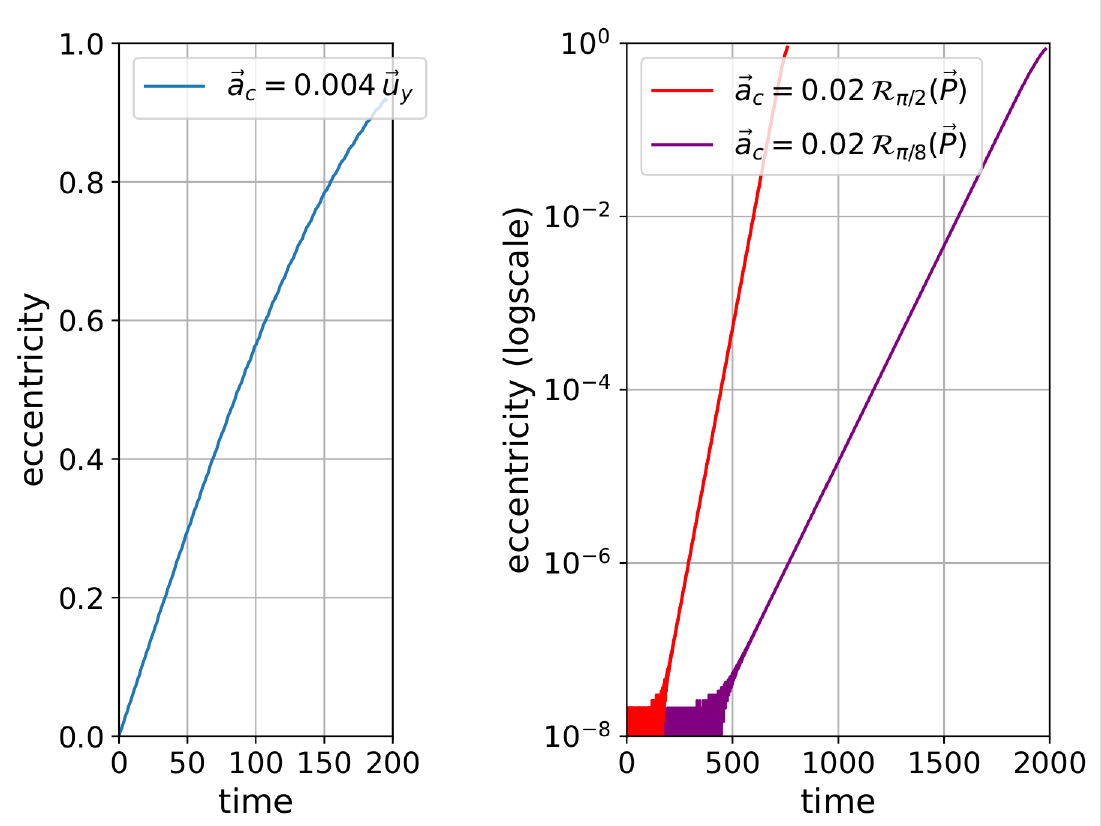}
    \caption{Orbital evolution of a particle feeling an acceleration $\vec{a}=-(1/r^2)\vec{u}_r+\vec{a}_c$. Top left panel\,: $\vec{a}_c=0.004\,\vec{u}_y$. Top right\,: $\vec{a}_c = 0.02\,\mathcal{R}_{\pi/2}(\vec{P})$. Bottom left panel\,: $\vec{a}_c = 0.02\,\mathcal{R}_{\pi/8}(\vec{P})$, where $\mathcal{R}_\phi$ is the rotation by an angle $\phi$ in the $x-y$ plane. Bottom right panel\,: evolution of the eccentricity in the three simulations (same color).
    }
    \label{fig:RK4}
  \end{center}
\end{figure}

This is confirmed by a simple numerical simulation where the motion of a test particle is integrated through a Runge-Kutta-4 algorithm with an acceleration given by $-r^{-2}\vec{u}_r+\vec{a}_c$ with $\vec{a}_c=0.004\vec{u}_y$ (units are chosen so that $G=M=1$ and the particle starts on a circular orbit of radius $1$).
The top left panel of \figref{fig:RK4} shows the resulting trajectory; it goes from circular to eccentric with a linear increase in the eccentricity (bottom right panel) and $\varpi=0$.

\subsection{Creating a feedback loop}

In a second simulation, we set the additional acceleration to be 
$\vec{a}_c = 0.02\cdot\mathcal{R}_{\pi/2}\left(\vec{P}\right)$, 
where $\mathcal{R}_{\phi}$ is a rotation of angle $\phi$ around the $z$-axis (in the $x-y$ plane).
Since the acceleration is now proportional to the eccentricity, this leads to an exponential growth of the eccentricity, as shown in the top right and bottom right panels of \figref{fig:RK4}. The initial orbit was again circular, but rounding errors are enough to initiate a positive feedback loop whereby the eccentricity triggers an acceleration which increases the eccentricity (as seen above). As a consequence, the Laplace vector grows exponentially in norm, keeping its direction constant (here $\vec{u}_x$).

If now the angle of the rotation is  not exactly $\pi/2$ but between $0$ and $\pi$, we observe a slower, but still exponential growth of the eccentricity, with a linear precession of the longitude of the pericentre. The result is shown in the bottom left and bottom right panels of \figref{fig:RK4} for an angle of $\pi/8$. This can be interpreted as $\vec{a}_c$ having a component perpendicular to $\vec{P}$, which yields the exponential growth of $e$, and a component parallel to $\vec{P}$, which yields the precession of the orbit.

\subsection{Link with the reflex instability}

The above results illustrate how a positive feedback loop can easily ignite in our set-up. The eccentricity growth will build an $m=1$ mode in the gas disc. In turn, this asymmetry in the density will accelerate the star, causing an indirect term proportional to the eccentricity.
For a simple eccentric gas disk (uniform eccentricity and longitude of periastron), the surface density perturbation due to the eccentric motion is symmetric with respect to major axis. Hence, the indirect term would be strictly parallel to the Laplace vector in this case. However, in \figref{fig:bary} the direction of the acceleration is not parallel to the star -- barycentre line. Therefore, it may have a component exciting the eccentricity and the $m=1$ mode in some part of the disc, yielding an exponential growth of the star -- barycentre distance, and a perpendicular component which makes the system precess linearly. Hence, the star describes a logarithmic spiral, as observed.

Describing analytically the response of a Keplerian gas disc to a uniform acceleration, and deriving the acceleration this response exerts on the star, are the missing ingredients to fully characterise the reflex instability. Nonetheless, this simple analysis in the two-body problem illustrates how the feedback loop may work.

\end{appendix}

% ================
% REFERENCES
%\bibliographystyle{aa}
\bibliographystyle{mnras}
\bibliography{./crida+.bib}
% ================

\end{document}